\documentclass[9pt,twocolumn,twoside]{pnas-new} 
\templatetype{pnasresearcharticle}
\usepackage{bm}
\usepackage{pdfpages}
\title{Asymmetric percolation drives a double transition in sexual contact networks}
\author[a]{Antoine Allard}
\author[b,c,d]{Benjamin M. Althouse} 
\author[e]{Samuel V. Scarpino}
\author[b,f,g,1]{Laurent H\'ebert-Dufresne}
\affil[a]{Centre de Recerca Matem\`atica, Edifici C, Campus Bellaterra, E-08193 Bellaterra (Barcelona), Spain}
\affil[b]{Institute for Disease Modeling, Bellevue, WA, 98005, USA}
\affil[c]{University of Washington, Seattle, WA, 98105, USA}
\affil[d]{New Mexico State University, Las Cruces, NM, 88003, USA}
\affil[e]{Department of Mathematics and Statistics and Complex Systems Center, University of Vermont, Burlington, VT, USA}
\affil[f]{Santa Fe Institute, Santa Fe, NM, 87501, USA}
\affil[g]{Department of Computer Science, University of Vermont, Burlington, VT, 05405, USA}
\leadauthor{Allard} 
\significancestatement{Zika virus (ZIKV) continues to be a threat to countries with conditions suitable for transmission, namely, adequate temperatures and the presence of competent mosquito vectors. Estimates of risk in other countries, based on the sexual transmission of ZIKV, may be underestimated due to inadequate surveillance. Here we formulate random network models of sexual transmission of ZIKV with asymmetric transmission (men being infectious for longer than women) and show that, contrary to previous work, there exists two epidemic thresholds and that certain men-who-have-sex-with-men communities could sustain transmission on their own. Our results also shed light on a new class of processes on random networks by providing a complete analysis of dynamics with multiple critical points.}

\authorcontributions{A.A. and L.H.-D. performed the calculations; all authors conceived of the project, interpreted the results, and produced the final manuscript.}
\authordeclaration{The authors declare no conflict of interest.}
\correspondingauthor{\textsuperscript{1}To whom correspondence should be addressed. E-mail: laurent@santafe.edu}


\begin{abstract}
Zika virus (ZIKV) exhibits unique transmission dynamics in that it is concurrently spread by a mosquito vector and through sexual contact. We show that this sexual component of ZIKV transmission induces novel processes on networks through the highly asymmetric durations of infectiousness between males and females -- it is estimated that males are infectious for periods up to ten times longer than females -- leading to an asymmetric percolation process on the network of sexual contacts. We exactly solve the properties of this asymmetric percolation on random sexual contact networks and show that this process exhibits two epidemic transitions corresponding to a core-periphery structure. This structure is not present in the underlying contact networks, which are not distinguishable from random networks, and emerges \textit{because} of the asymmetric percolation. We provide an exact analytical description of this double transition and discuss the implications of our results in the context of ZIKV epidemics. Most importantly, our study suggests a bias in our current ZIKV surveillance as the community most at risk is also one of the least likely to get tested.
\end{abstract}
\dates{This manuscript was compiled on \today}
\begin{document}
%
\verticaladjustment{-2pt}
\maketitle
\thispagestyle{firststyle}
\ifthenelse{\boolean{shortarticle}}{\ifthenelse{\boolean{singlecolumn}}{\abscontentformatted}{\abscontent}}{}
\dropcap{A}bstract modeling of epidemics on networks remains an active field because some of the most basic features of epidemics are still misunderstood. The classic model is quite simple \cite{Newman2002}: disease spreads stochastically, with a fixed transmission probability, $T$, through contacts around a given patient zero. The outbreak dies quickly if $T$ is too small, but spreads to a macroscopic fraction $S$ of the entire population if $T$ is larger than a threshold $T_\mathrm{c}$. At $T_\mathrm{c}$, most of the typical insights from phase transition theory are valuable. For instance, the sizes of microscopic outbreaks follow a power-law distribution such that the expected size of microscopic outbreaks, $\langle s \rangle$, indicates the position of a phase transition. Indeed, as $T$ increases, $\langle s \rangle$ monotonically increases, diverges exactly at $T_\mathrm{c}$, and then monotonically goes down; meanwhile the expected macroscopic epidemic size, $S$, starts increasing monotonically at $T_\mathrm{c}$.


However, simple modifications to this model can dramatically alter its phenomenology. The epidemic threshold can vanish in networks with a scale-free degree distribution \cite{Pastor-Satorras2001a} or in growing networks \cite{Althouse2014}. The phase transition can be discontinuous in the case of complex contagions with threshold exposition or reinforcement \cite{Dodds2004}, interacting epidemics \cite{Hebert-Dufresne2015,Cai2015}, or adaptive networks \cite{Gross2006,Marceau2010,Scarpino2016}. Recently, a unique phenomenon of double phase transitions has also been observed numerically when networks have a very heterogeneous and clustered structure \cite{Colomer-de-Simon2014,Bhat2017}.

\begin{figure*}
  \centering
  \includegraphics[width=0.49\linewidth]{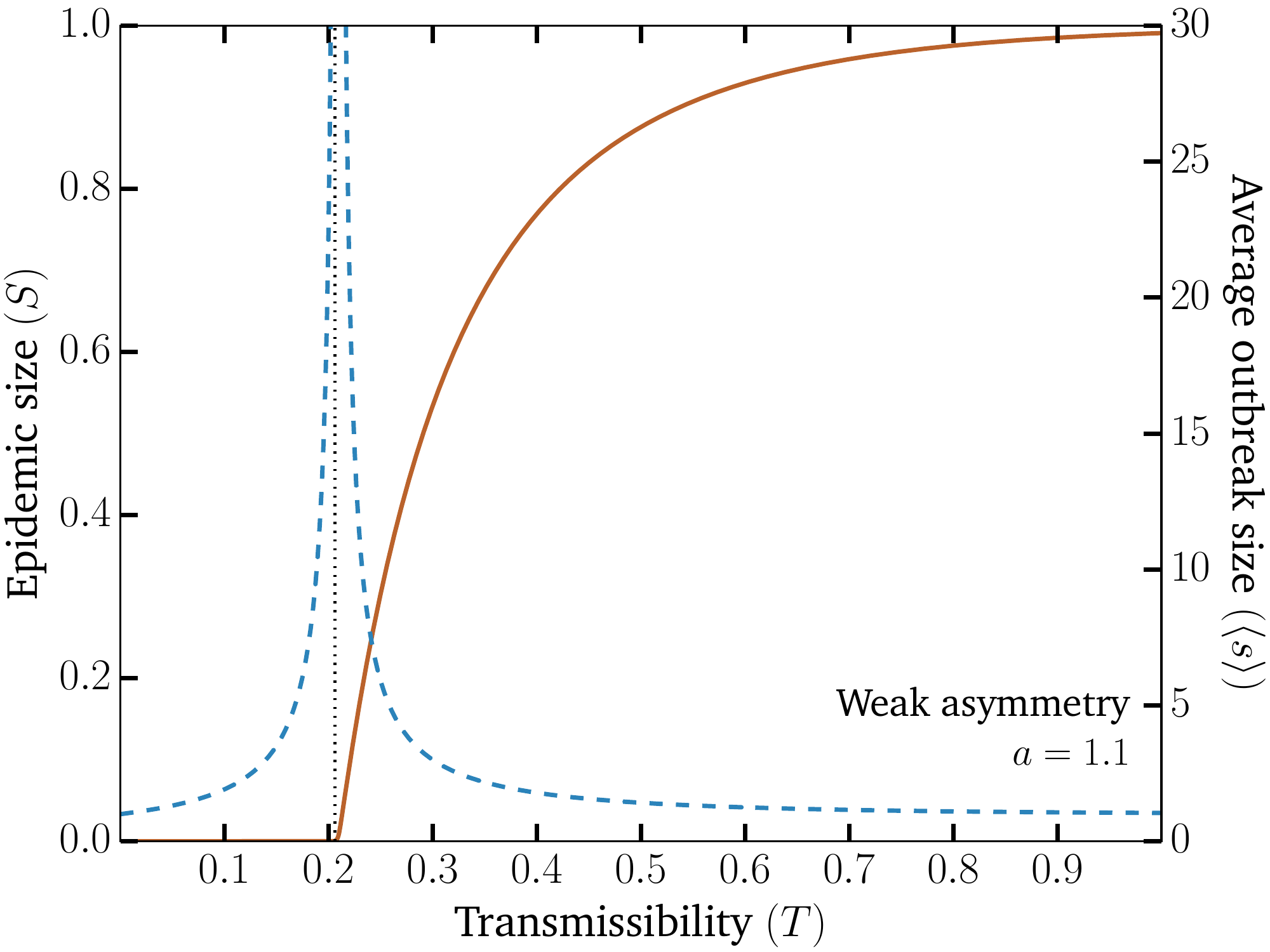}
  \includegraphics[width=0.49\linewidth]{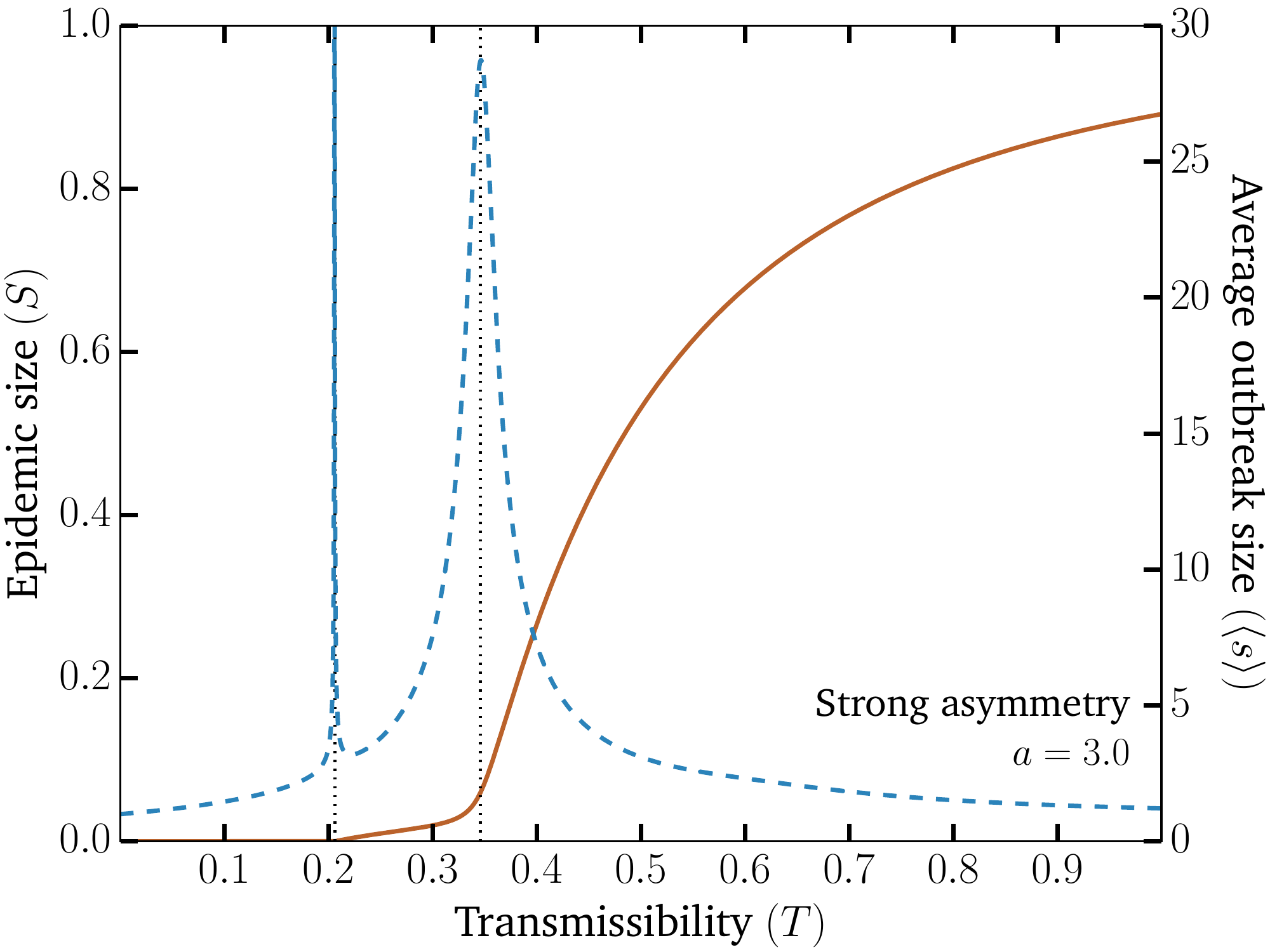}
  \caption{\label{fig:notamovie}\textbf{Emergence of the second transition as asymmetry increases.} The solid lines show the expected fraction of the population in the extensive component ($S$, left axis). The dashed lines show the average size of small, non-extensive components ($\langle s \rangle$, right axis). The divergence of the average size of small components marks the phase transition after which the extensive components grows with the transmission probability $T$. The vertical dotted black lines show the thresholds. (left) With a small asymmetry between transmission values $\{T_{ij}\}$ as a function of node types, we recover the classic epidemic transition. (right) With a larger asymmetry, a second peak in the average size of small components appears. The first, $T_\mathrm{c}^{(1)}$, corresponds to the global epidemic threshold of the population. The second, $T_\mathrm{c}^{(2)}$, corresponds to the invasion of the large heterosexual subpopulation. The threshold $T_\mathrm{c}^{(1)}$ corresponds to the value of $T$ such that the largest eigenvalue of the Jacobian matrix of \eqref{eq:self_consistency} equals 1. The second threshold $T_\mathrm{c}^{(2)}$ is obtained similarly but with the probability of transmission between homosexual males set to zero. The homo-/bi-/heterosexual subpopulations represent 5\%, 3\% and 92\%, respectively and are equally split between males and females. The degrees are distributed according to a Poisson distribution, $p_k = e^{-\langle k \rangle} \langle k \rangle^k/k!$, with an average degree, $\langle k \rangle$, equal to 5. See Supporting Information for further details.}
\end{figure*}

\begin{figure}[t!]
  \centering
  \includegraphics[width=0.98\linewidth]{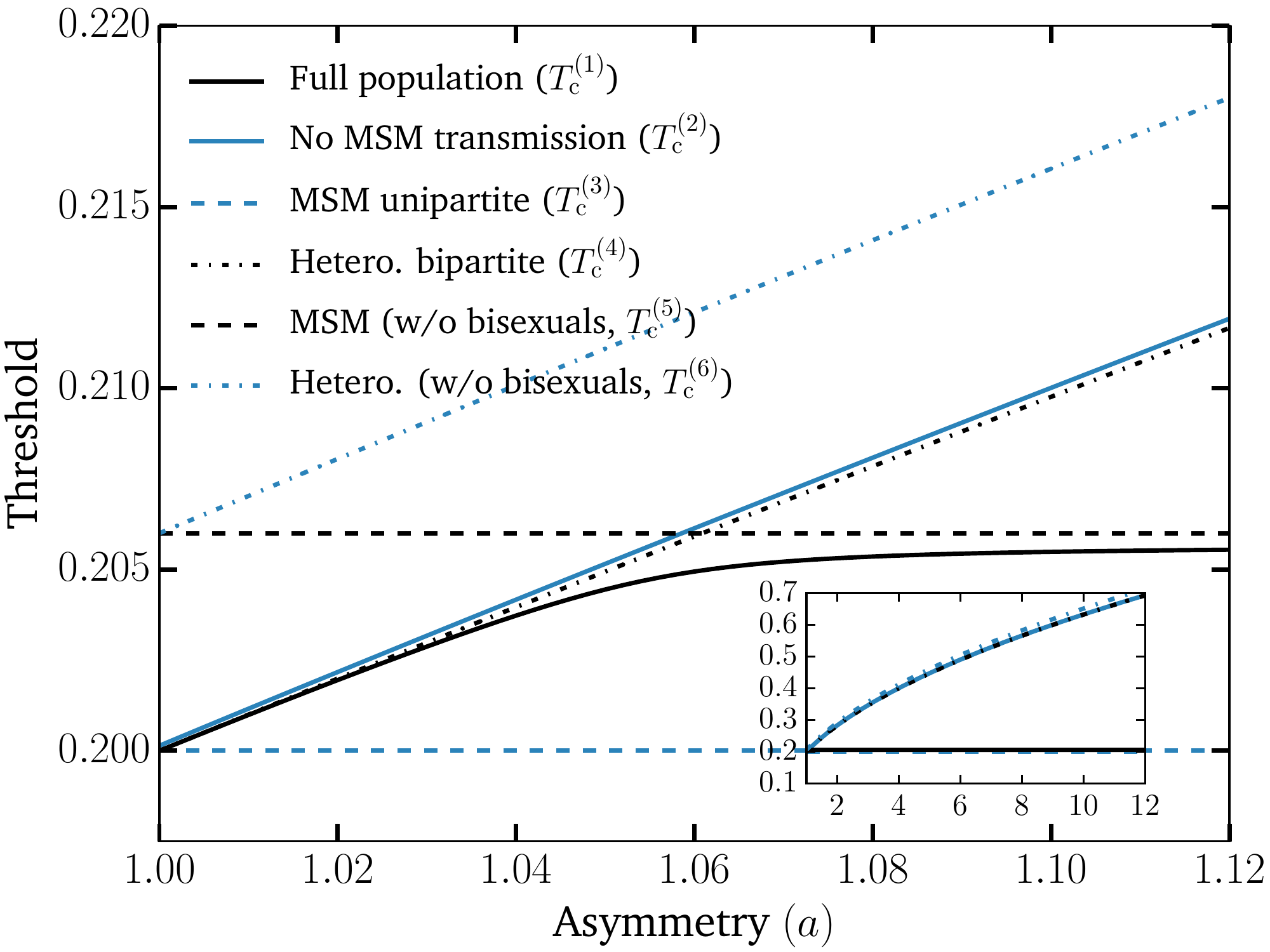}
  \caption{\textbf{Separation of thresholds with increasing asymmetry.} We show the two thresholds (critical points) $T_\mathrm{c}^{(1)}$ and $T_\mathrm{c}^{(2)}$ discussed in the main text (solid lines), as well as the thresholds for different subpopulations (dotted lines) which can be easily calculated and further support our interpretation of the phenomenology. The first threshold, $T_\mathrm{c}^{(1)}$, corresponds to the epidemic threshold for the full population. The second threshold, $T_\mathrm{c}^{(2)}$, is computed by setting the transmission between men-having-sex-with-men (MSM) to zero. The ``MSM unipartite'' and ``Hetero. bipartite'' lines show the epidemic threshold should the network be only populated with MSM or heterosexuals, respectively. They are defined as $\langle k \rangle_\mathrm{e} T_\mathrm{c}^{(3)} = 1$ and $[\langle k \rangle_\mathrm{e} T_\mathrm{c}^{(4)}]^2/a = 1$, where $\langle k \rangle_\mathrm{e} = \langle k(k-1) \rangle / \langle k \rangle$ is the average \textit{excess} degree of the nodes \cite{Newman2001}. The last two thresholds correspond to the contributions to $T_\mathrm{c}^{(1)}$ and $T_\mathrm{c}^{(2)}$ that involve exclusively the homosexual male or heterosexual subpopulations, respectively. They are the solutions of $\alpha_{0|0} \langle k \rangle_\mathrm{e} T_\mathrm{c}^{(5)} = 1$ and $ \alpha_{4|5}\alpha_{5|4}[\langle k \rangle_\mathrm{e} T_\mathrm{c}^{(6)}]^2/a = 1$, where nodes of type 0, 4 and 5 correspond to homosexual males, heterosexual males and heterosexual females, respectively. These results therefore support the interpretation that the first threshold corresponds to the invasion of the MSM subpopulation (with $T_\mathrm{c}^{(3)}$ and $T_\mathrm{c}^{(5)}$ acting as lower and upper bounds), and that the second threshold is due to the invasion of the remaining population (with $T_\mathrm{c}^{(4)}$ and $T_\mathrm{c}^{(6)}$ acting as lower and upper bounds). The inset shows the growing separation of the two main thresholds as asymmetry increases to values close to what we expect for ZIKV. The same parameters as in Fig.~\ref{fig:notamovie} were used.}
  \label{fig:forking}
\end{figure}

The current Zika virus (ZIKV) epidemic exhibits two unique properties. First, while the main transmission pathway for ZIKV is through a mosquito vector (predominantly {\em Aedes aegypti} or {\em Aedes albopictus}~\cite{Althouse2016, Althouse2015}), a feature which has its own type of well-studied model and behavior \cite{Smith2012,Althouse2012,Althouse2015}, it can also spread through sexual contacts~\cite{Althaus2016,Yakob2016}. Second, the probability of sexual transmission is highly asymmetric between males and females. Although this is also true for other sexually transmitted infections such as HIV~\cite{Padian1991}, it reaches an extreme	 level of asymmetry in the case of ZIKV. Indeed, males can be infectious for over 180 days~\cite{Nicastri2016} while females are infectious for less than 20 days~\cite{Prisant2016}. Assuming a symmetric risk of transmission per contact, males would be 10 times more likely to transmit to a partner than females. This is, however, a rather conservative estimate since male-to-female transmissions tend to be more likely than the opposite \cite{Padian1991,Boily2009}.

The dynamics of the ZIKV epidemic is well understood in countries where the vector-borne pathway dominates \cite{Zhang2017}. However, with travelers moving to and from endemic regions, the potential of ZIKV as an emerging STI in regions without the mosquito vector remains to be fully assessed. Indeed, with only few reported cases of sexual transmission of ZIKV -- including male-to-male, male-to-female, and female-to-male \cite{Yakob2016} -- the scientific community still struggles to reach a consensus on the impact of sexual transmission of ZIKV~\cite{Folkers2017, moreira2017sexually}. It is therefore imperative to investigate the extent to which canonical knowledge about emerging infectious diseases applies to the threat assessment of ZIKV as an STI.

We model the ZIKV sexual transmission through asymmetric percolation on random sexual contact networks and solve it exactly using a multitype (multivariate) generating function formalism \cite{Allard2015}. We then show how the asymmetric percolation leads to a double transition. Interestingly, the formulation of our model allows us to provide a first analytic framework for the aforementioned numerical results on double transitions. More importantly, this allows us to identify two different thresholds for ZIKV to be endemic as an STI in regions where the mosquito vector is absent, but where travellers to/from endemic regions can spark a sexual epidemic when they return/visit. We also find that, in the large interval of parameter space between those two thresholds, the asymmetric percolation creates a core-periphery structure in a system where there was none. Finally, we discuss the implications of this core-periphery structure for the surveillance and control of the ZIKV epidemic, and provide policy guidelines.

\section*{Results}
Inspired by the sexual transmission of ZIKV, we investigate the effect of asymmetry on bond percolation on networks, and show that it yields outcomes akin to the double phase transitions observed numerically in other contexts \cite{Colomer-de-Simon2014,Bhat2017}. To isolate the effect of asymmetry alone and thus provide a clear proof of concept, we consider a very simple model in which nodes belong to one of 6 types based on their sex and sexual orientation (i.e., female/male and homo-/bi-/heterosexual). Each node is assigned a number of contacts, $k$, independently of its type (i.e., all nodes have the same degree distribution $\{p_k\}_{k\geq0}$), and links are created randomly via a simple stub-matching scheme constrained by the sexual orientations \cite{Newman2001,Allard2015}. For instance, bisexual males choose their partners randomly in the pools of heterosexual females, bisexual males and females, and homosexual males. This implies that there is no correlation between the type of a node and its number of contacts, and consequently no core-periphery structure. In fact, this model generates well-mixed contact networks that are indistinguishable from networks generated with the configuration model and the same degree distribution (see Supporting Information).

\begin{figure*}
  \centering
  \includegraphics[width=0.45\linewidth]{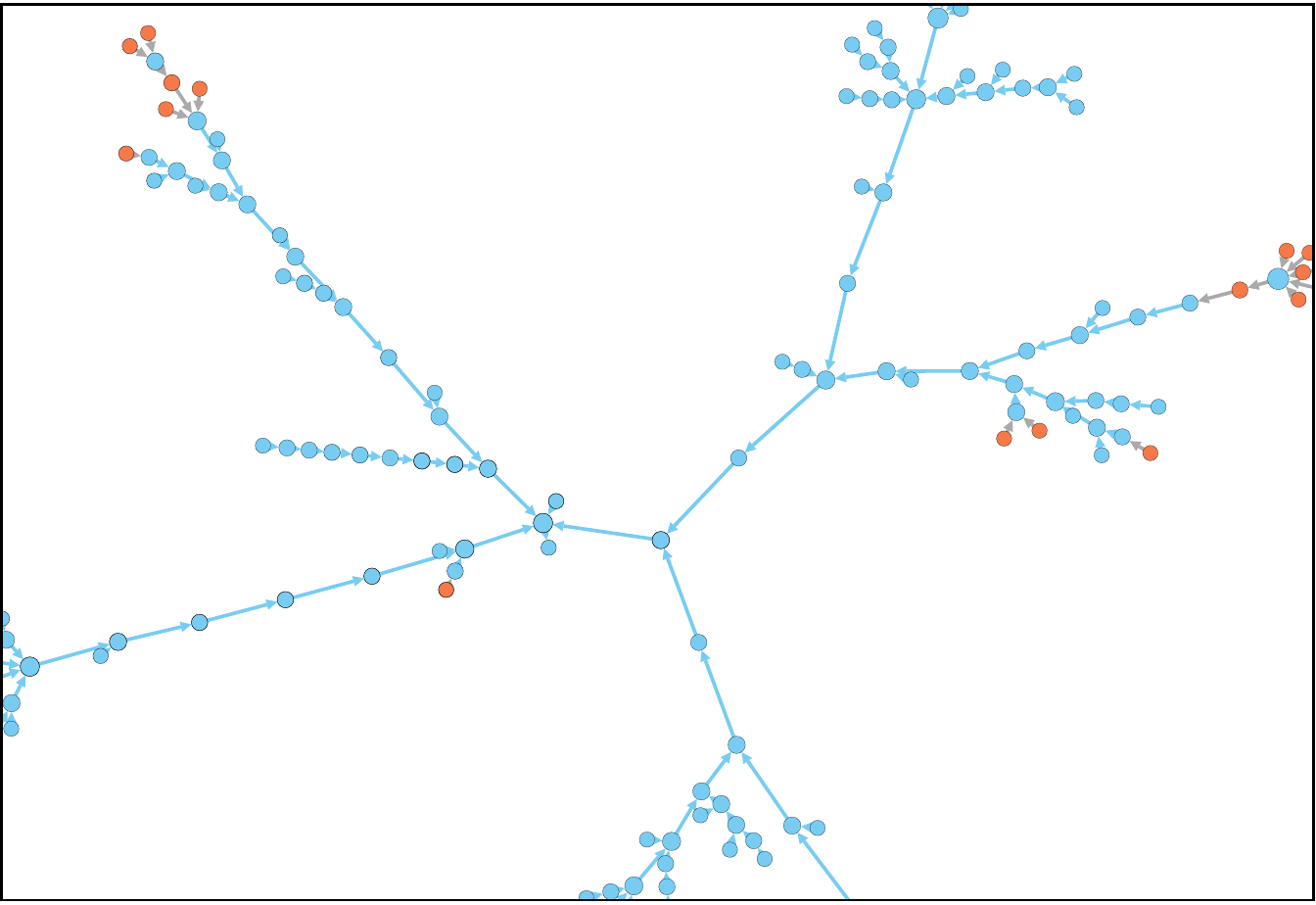}
  \hspace{0.025\linewidth}
  \includegraphics[width=0.45\linewidth]{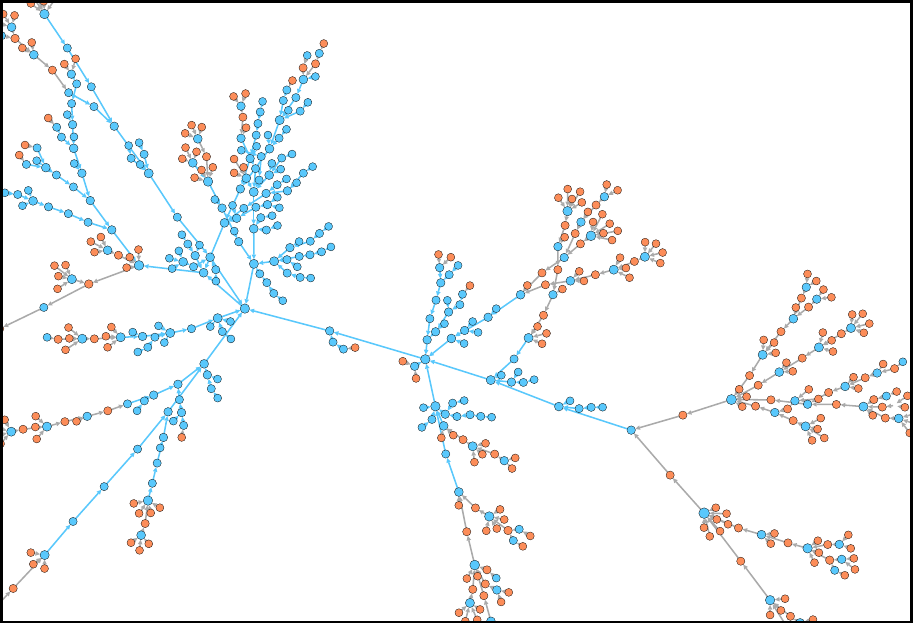}
  \caption{\label{fig:networks}\textbf{Composition of the components as the transmissibility increases.} Nodes corresponding to males and females are shown in blue and orange, respectively, and arrows indicate \textit{who infected whom}. The same parameters as for Fig.~\ref{fig:notamovie} have been used with asymmetry $a = 10$. (left) At $T_\mathrm{c}^{(1)}<T=0.45<T_\mathrm{c}^{(2)}$, the infection mostly follows the MSM sub-population with minimal and sub-critical spillovers in the remaining population. (right) At $T=T_\mathrm{c}^{(2)} \simeq 0.632$, the spillover causes cascades of power-law distributed sizes into the heterosexual population.}
\end{figure*}

While these networks are originally undirected, asymmetric percolation implies that links can be more likely to exist (i.e., transmit) in one direction than in the other, thus inducing an effective semi-directed structure to the networks \cite{Allard2009}. In other words, $T_{ij} \neq T_{ji}$ in general, with $T_{ij}$ being the probability of transmission from a node of type $i$ to a node of type $j$ (hereafter we denote $\mathcal{N}$ the set of the six possible types of nodes). In particular, we set $T_{ij}=T$ for every $i, j \in \mathcal{N}$ except when $i$ corresponds to a female, in which case we set $T_{ij}=T/a$ to enforce asymmetric probabilities of transmission (i.e., females are $a$ times \textit{less}  likely to transmit ZIKV than males).

We adapt the formalism presented in Ref.~\cite{Allard2015} to compute the epidemic threshold and the expected final size of outbreaks in the limit of large networks. It is worth pointing out that since asymmetric percolation (i.e., whenever $a \neq 1$) induces an effective semi-directed structure to the networks, the probability for the existence of an extensive connected component (i.e, an epidemic) does not equal to its relative size as for symmetric, traditional bond percolation (i.e., $a = 1$). Here we focus on the relative size for the sake of conciseness; we refer the readers to the Supporting Information for full details of the analysis and numerical validation.

To obtain the relative size of the extensive component, we define $v_i$ as the probability that a neighbor of type $i$ is not in the extensive component, which we solve by a self-consistent argument. If the neighbor of a node is not in the extensive component, then none of its other neighbors should be in it either. The probability that the neighbor has a degree equal to $k$ being $kp_k/\langle k \rangle$, with $\langle k \rangle = \sum_k kp_k$, this self-consistent argument can be written as
\begin{align} \label{eq:self_consistency}
  v_i & = \sum_{k} \frac{k p_k}{\langle k \rangle} \left[ \sum_{j \in \mathcal{N}} \alpha_{j|i} (1 - T_{ji} + T_{ji}v_j) \right]^{k-1} \; .
\end{align}
where $\alpha_{j|i}$ is the probability that a neighbor of a node of type $i$ is of type $j$ (i.e., $\sum_{j}\alpha_{j|i}=1$ for any $i$). Solving this equation for every $i\in\mathcal{N}$, the probability that a node of type $i$ is part of the extensive component, $S_i$, corresponds to the probability that at least one of its neighbors is in it as well
\begin{align}
  S_i & = 1 - \sum_{k} p_k \left[ \sum_{j \in \mathcal{N}} \alpha_{j|i} (1 - T_{ji} + T_{ji}v_j) \right]^k \; .
\end{align}
The relative size of the extensive component is then $S = \sum_{i\in\mathcal{N}} w_i S_i$, where $w_i$ is the fraction of the nodes that are of type $i$. Below the epidemic or percolation threshold, every $v_i$ is equal to 1 since there is no extensive component. The percolation threshold corresponds to the point where the largest eigenvalue of the Jacobian matrix of \eqref{eq:self_consistency} equals 1.

The distribution of the composition of the small, non-extensive components can be computed in a similar fashion (see Supporting Information for full details). Let us define the probability generating function (pgf) $H_i(\bm{x})$ whose coefficients correspond to the probability that a neighbor of type $i$ leads to a small component of a given composition (i.e., the number of nodes of type $j$ is given by the exponent of $x_j$). Invoking the same self-consistency argument as above, the pgfs are the solution of 
\begin{align}
  H_i(\bm{x}) & = x_i \sum_{k} \frac{k p_k}{\langle k \rangle} \left[ \sum_{j \in \mathcal{N}} \alpha_{j|i} [1 - T_{ij} + T_{ij}H_j(\bm{x})] \right]^{k-1} \; ,
\end{align}
where the extra $x_i$ has been added to account for the neighbor of type $i$ itself. Similarly, the small component that can be reached from a node of type $i$ is therefore given by
\begin{align} \label{eq:small_component}
  K_i(\bm{x}) & = x_i \sum_{k} p_k \left[ \sum_{j \in \mathcal{N}} \alpha_{j|i} [1 - T_{ij} + T_{ij}H_j(\bm{x})] \right]^{k} \; .
\end{align}
The distribution of the composition of the small components is $K(\bm{x}) = \sum_{i\in\mathcal{N}} w_i K_i(\bm{x})$. It is worth noting that whenever $S>0$, the distribution generated by $K(\bm{x})$ is no longer normalized, $K(\bm{1})<1$, such that the average number of nodes of type $i$ in the small components is
\begin{align}
  \langle s_i \rangle = \frac{1}{K(\bm{1})} \left. \frac{dK(\bm{\bm{x}})}{dx_i}\right|_{\bm{x}=\bm{1}} \ .
\end{align}

\begin{figure*}[t!]
  \centering
  \includegraphics[width=0.49\linewidth]{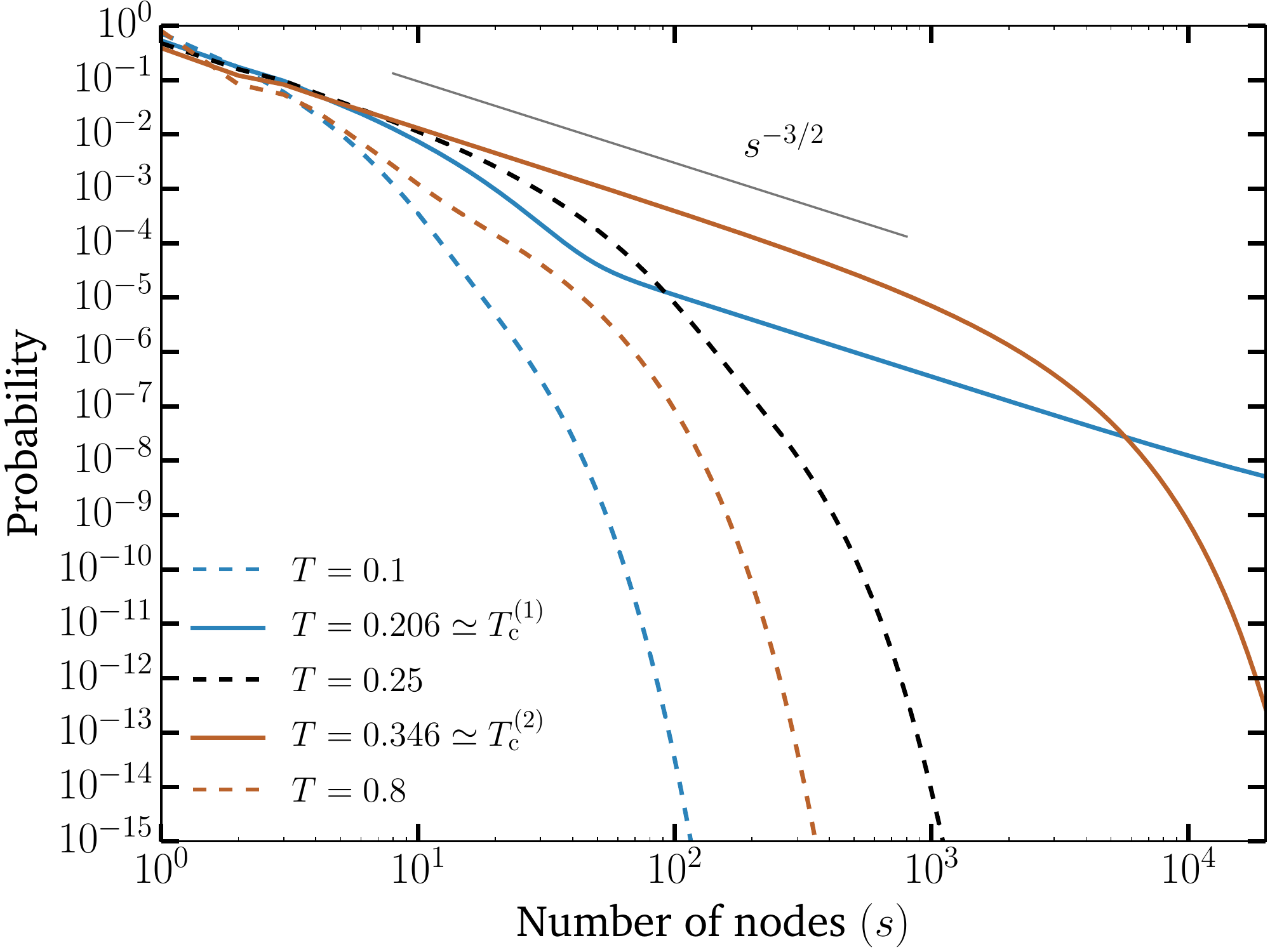}
  \includegraphics[width=0.49\linewidth]{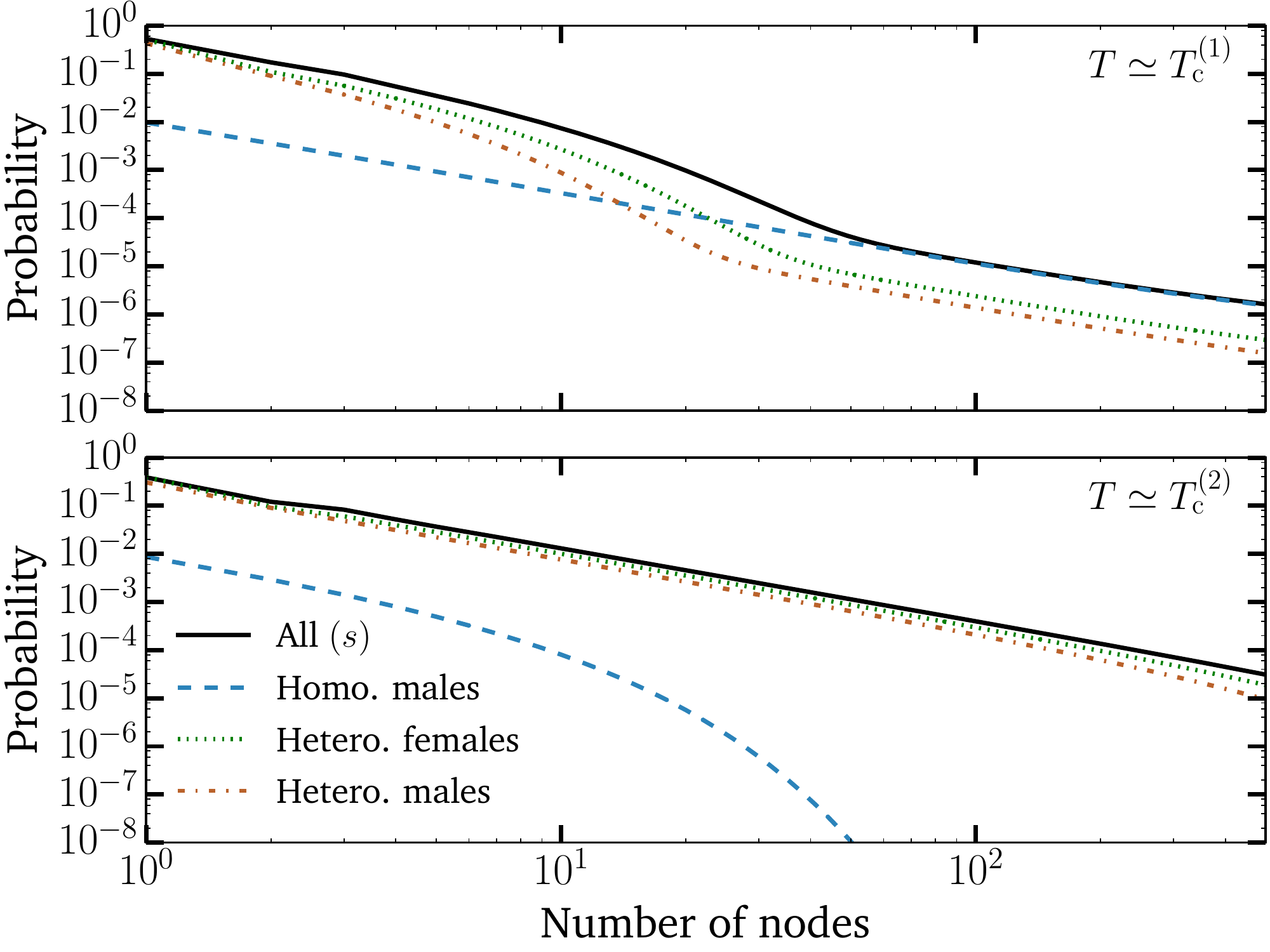}
  \caption{\textbf{Distribution of the size and composition of small components.} (left) We find power-law scaling of small outbreak sizes with scaling exponent $-3/2$, as expected from classic phase transition theory \cite{Newman2001}, at both $T_\mathrm{c}^{(1)}$ and $T_\mathrm{c}^{(2)}$. However, unlike classic phase transitions, only the tail of the distribution follows a power-law at $T_\mathrm{c}^{(1)}$, while at $T_\mathrm{c}^{(2)}$ we find a robust power-law over many orders of magnitude before the distribution falls with the expected exponential cut-off. This cut-off goes to infinity when the size of the MSM community goes to zero, in which case $T_\mathrm{c}^{(2)}$ now becomes the prominent critical point. Notice that the size of the components goes back to an homogeneous distribution in-between the two epidemic thresholds. (upper right) At $T_\mathrm{c}^{(1)}$, the power-law tail in the component size distribution is mainly due to the critical core of homosexual males while the exponential behavior is mainly due to heterosexuals. The power-law tail in the distributions of heterosexuals is due to spillovers from the critical core. (lower right)  At $T_\mathrm{c}^{(2)}$, the power-law portion of the distribution is due to heterosexuals now forming a critical core while homosexual males, being already almost exclusively in the extensive component, do not contribute. All curves were obtained by solving \eqref{eq:small_component} with asymmetry $a=3$ and the parameters given in the caption of Fig.~\ref{fig:notamovie}.}
  \label{fig:small_component_size}
\end{figure*}

An example of the general phenomenology is shown in Fig.~\ref{fig:notamovie}. Unlike the classic epidemic transition picture, where $\langle s \rangle$ diverges at the epidemic threshold where the macroscopic epidemic emerges, we now find two peaks in $\langle s \rangle$. This double transition is similar to numerical results from Ref.~\cite{Colomer-de-Simon2014}, but here observed without the need for either strong clustering nor heterogeneity in degree distribution. In fact, we used the homogeneous Poisson degree to ensure that the asymmetry in the transmission is the only salient feature of the model. Interestingly, as shown in Fig.~\ref{fig:forking}, $T_\mathrm{c}^{(1)}$ and $T_\mathrm{c}^{(2)}$ are virtually equal for small values of the asymmetry. As asymmetry increases, the peak separates thus yielding a double transition corresponding to an effective core-periphery organization in the network of infections. The core then corresponds to the men-having-sex-with-men (MSM) population where infections are more frequent than in the remaining population. Figure~\ref{fig:networks} shows the network of \textit{who infected whom} for two values of $T$. For $T_\mathrm{c}^{(1)} < T < T_\mathrm{c}^{(2)}$, the extensive component is mostly composed of one type of nodes and any spillover in the other types quickly dies out. However, at $T=T_\mathrm{c}^{(2)}$, these spillovers now cause cascades into other types with truncated power-law distributed sizes (see Fig.~\ref{fig:small_component_size}). For $T>T_\mathrm{c}^{(2)}$, the extensive component recovers the well-mixed structure of the original underlying network.

Altogether, the second peak in the average size of outbreaks, $\langle s \rangle$, corresponds to a transition between subcritical and supercritical spillover in a less susceptible sub-population, but not to a second phase transition in the classic sense. Indeed, the analytical nature of our results allows confirm the null critical exponent observed in Ref.~\cite{Colomer-de-Simon2014} for the scaling of the height of the second susceptibility peak with regards to system size. Even in the infinite system considered by our calculations, the peak saturates, which is the only possible outcome for a system whose order parameter is already non-zero. Interestingly, a critical power-law-like behavior is nonetheless observed in the heterosexual population at both thresholds. Moreover, our results suggest that the asymmetry in transmission probability is reflected in the asymmetric prevalence within the male and female heterosexual populations, which is reminiscent of recent empirical results~\cite{coelho2016higher}.

Based on our results, we can summarize the phase diagram of the ZIKV epidemic in 3 possible outcomes. First, with $T < T_\mathrm{c}^{(1)}$, all outbreaks are microscopic, quickly die out, and mostly infect MSM. Second, with $T_\mathrm{c}^{(1)} < T < T_\mathrm{c}^{(2)}$, we now see a macroscopic epidemic within the network of homosexual contacts between males, with microscopic spill-over into the rest of the population via bisexual males. Third, with $T > T_\mathrm{c}^{(2)}$, we now find a more classic epidemic scenario in the sense that it is of macroscopic scale in most of the population. It is also worth mentioning that this phenomenology is robust to the presence of multiple infectious seeds sparking outbreaks (see Supporting Information). Our results are thus valid beyond ZIKV for any infections with asymmetry in probabilities of direct transmission, regardless of whether or not there is also vector transmission.
%
%
%
\section*{Discussion}
We developed a network model of ZIKV transmission highlighting the importance of asymmetric sexual transmission between males and females. We find a double transition generated by a core group of MSM that could maintain ZIKV transmission without the presence of a viable mosquito vector, such as in regions where people may have brought back ZIKV with them after a trip to endemic regions. These results are unique as previous models showing double transitions relied on the need for strong clustering and heterogeneity in degree distribution.

Our study carries important consequences for the ongoing ZIKV epidemic and stresses the large knowledge gap in the sexual transmission of ZIKV~\cite{moreira2017sexually}. The aim of our work is to present the epidemiological consequences of possible sustained sexual transmission. While there are many unknowns, recent work demonstrates a) multiple anecdotal cases of sexual transmission of ZIKV in humans~\cite{moreira2017sexually, russell2016male,foy2011probable,d2016evidence}, b) multiple separate animal models demonstrating sexual transmission~\cite{morrison2017animal,duggal2017frequent,yockey2016vaginal}, c) strong asymmetries between durations of ZIKV shedding in semen and vaginal secretions~\cite{Nicastri2016,Prisant2016}, and d) differential risk between sexes for ZIKV infection in sexually active populations. Indeed, recent work has identified 90\% more ZIKV infections in women between 15 and 65 than men of the same age in Rio de Janero~\cite{coelho2016higher} adjusted for gender-related health-seeking behavior and pregnancy status. Importantly, this risk difference was not seen in women less than 15 years of age or greater than 65, indicating the potentially large impact of sexual transmission of ZIKV in a country with known ongoing vectored transmission of ZIKV. A similar situation has also been observed in Colombia \cite{PAHOZikaColombia} and in the Dominican Republic \cite{PAHOZikaDominicanRepublic}. While more research on the epidemiological impacts and basic biology of sexual ZIKV transmission is needed, there is compelling need to be prepared with epidemiological studies examining transmission on a population-scale.

We demonstrated that potential ZIKV persistence in MSM, even if barely critical within that sub-population, could cause subcritical but dramatic spillover into the heterosexual community.  ZIKV infections in adults are largely asymptomatic \cite{Duffy2009} and, therefore, most testing occurs in the roughly $20\%$ of cases that are symptomatic or in individuals seeking to have children \cite{Lessler2016}. The vast majority of these individuals will be outside of the MSM community \cite{Lessler2016}. This means that the community most at risk is also one of the least likely to get tested. To avoid underestimating the spread of ZIKV, it is therefore important for health officials and policy makers to keep its unique behavior and phenomenology in mind.

Given the extent of foreign travel to locations endemic with ZIKV, public health practitioners should be aware of the potential for infectious introduction into local MSM communities. Travel history as well as sexual history should be employed when evaluating an occult fever. Cities which have a viable vector for ZIKV should be doubly aware of the potential transmission routes of ZIKV. As it stands, current estimates of the basic reproductive number, $R_0$, of ZIKV may be too low as they fail to account for sustained sexual transmission~\cite{Allard2016,Althaus2016,Yakob2016,Miller2017}. Important future work will be to accurately estimate $R_0$ of ZIKV across various settings with differing sexual practices and mosquito fauna.

\acknow{A.A. acknowledges support from the Fonds de recherche du Qu\'ebec -- Nature et technologies. L.H.-D. acknowledges the Santa Fe Institute, the James S. McDonnell Foundation Postdoctoral Fellowship and the National Science Foundation under Grant DMS-1622390. B.M.A. and L.H.-D. thank Bill and Melinda Gates for their support of this work and their sponsorship through the Global Good Fund. The funders had no role in study design, data collection and analysis, decision to publish, or preparation of the manuscript.}
\showacknow

\pnasbreak

\pnasbreak

\clearpage


\section*{Supplementary material}

# Asymmetric percolation drives a double transition in sexual contact networks
# — Supporting Information —

Antoine Allard,[1] Benjamin M. Althouse,[2,3,4] Samuel V. Scarpino,[5] and Laurent Hébert-Dufresne[2,6,7]

[1]*Centre de Recerca Matemàtica, Edifici C, Campus Bellaterra, E-08193 Bellaterra (Barcelona), Spain*
[2]*Institute for Disease Modeling, Bellevue, WA, 98005, USA*
[3]*University of Washington, Seattle, WA, 98105, USA*
[4]*New Mexico State University, Las Cruces, NM, 88003, USA*
[5]*Department of Mathematics and Statistics and Complex Systems Center, University of Vermont, Burlington, VT, USA*
[6]*Santa Fe Institute, Santa Fe, NM, 87501, USA*
[7]*Department of Computer Science, University of Vermont, Burlington, VT, 05405, USA*
(Dated: June 27, 2017)

The analytical approach used in the main text is an adaptation of a more general formalism presented in Ref. [1]. We derive in more details its equations, validate its predictions with numerical simulations, and demonstrate some of the claims made in the main text. We also provide further details about the parameters used in the calculations. Finally, the general phenomenology and the results presented in the main text are further validated with simulations using realistic data.

The formalism presented in Ref. [1] is a generalization of the configuration model [2] in which nodes and connections are distinguished via types of nodes and stubs (i.e., half-links that are connected to create motifs). We consider a case in which there are 6 types of nodes based on the sex and sexual orientation of individuals (i.e., female/male and homo-/bi-/heterosexual). We denote the set of such types $\mathcal{N}$ and say that a fraction $w_i$ of all nodes are of type $i$. Each node is assigned a number of contacts, $k$, independently of its type (i.e., all nodes have the same degree distribution $\{p_k\}_{k \geq 0}$), and links are created by randomly matching stubs respecting sexual orientations. Mathematically, this is encoded in the conditional probabilities $\{\alpha_{j|i}\}_{i,j \in \mathcal{N}}$ giving the probability that a link leaving a node of type $i$ leads to a node of type $j$ (see Sec. C for their numerical values). These probabilities are subjected to the constraints

$$\sum_{l \in \mathcal{N}} \alpha_{l|i} = 1 \; ; \qquad \alpha_{j|i} w_i = \alpha_{i|j} w_j \; , \tag{S1}$$

for every $i, j \in \mathcal{N}$. The latter constraint is due to the undirected nature of links and implies that there are as much links running from nodes of type $i$ towards nodes of type $j$ than in the opposite direction. Note that since nodes are only connected via links, the set of types of stubs is the same as the set of nodes and need not be considered explicitly (see Sec. IV of Ref. [1]).

Because links are created randomly, the probability that a node of type $i$ and degree $k$ has a set of $\boldsymbol{k} := \{k_j\}_{j \in \mathcal{N}}$ neighbors of each type is simply

$$P(\boldsymbol{k}|i, k) := \frac{k!}{\prod_{j \in \mathcal{N}} k_j!} \prod_{j \in \mathcal{N}} \left[ \alpha_{j|i} \right]^{k_j} \tag{S2}$$

with the constraint that $\sum_{j \in \mathcal{N}} k_j = k$. Given that links running from nodes of type $i$ to nodes of type $j$ are independently removed with probability $1 - T_{ij}$, the probability that a node of type $i$ and original degree $k$ has $\boldsymbol{m}$ neighbors is

$$P(\boldsymbol{m}|i, k) := \sum_{\boldsymbol{k} \geq \boldsymbol{m}} P(\boldsymbol{k}|i, k) \prod_{j \in \mathcal{N}} \binom{k_j}{m_j} T_{ij}^{m_j} (1 - T_{ij})^{k_j - m_j} \; , \tag{S3}$$

where the summation runs over $k_j \geq m_j$ for every $j \in \mathcal{N}$. The multivariate generating function associated with the joint degree distribution for nodes of type $i$ is

$$\begin{aligned}
g_i(\boldsymbol{x}) &:= \sum_k p_k \sum_{\boldsymbol{m}} P(\boldsymbol{m}|i,k) \prod_{j \in \mathcal{N}} x_j^{m_j} \\
&= \sum_k p_k \sum_{\boldsymbol{m}} \sum_{\boldsymbol{k} \geq \boldsymbol{m}} P(\boldsymbol{k}|i,k) \prod_{j \in \mathcal{N}} \binom{k_j}{m_j} [T_{ij} x_j]^{m_j} (1-T_{ij})^{k_j - m_j} \\
&= \sum_k p_k \sum_{\boldsymbol{k}} P(\boldsymbol{k}|i,k) \prod_{j \in \mathcal{N}} \sum_{m_j=0}^{k_j} \binom{k_j}{m_j} [T_{ij} x_j]^{m_j} (1-T_{ij})^{k_j - m_j} \\
&= \sum_k p_k \sum_{\boldsymbol{k}} \frac{k!}{\prod_{j \in \mathcal{N}} k_j!} \prod_{j \in \mathcal{N}} [\alpha_{j|i}]^{k_j} \sum_{m_j=0}^{k_j} \binom{k_j}{m_j} [T_{ij} x_j]^{m_j} (1-T_{ij})^{k_j - m_j} \\
&= \sum_k p_k \sum_{\boldsymbol{k}} \frac{k!}{\prod_{j \in \mathcal{N}} k_j!} \prod_{j \in \mathcal{N}} [\alpha_{j|i}]^{k_j} [1 - T_{ij} + T_{ij} x_j]^{k_j} \\
&= \sum_k p_k \left[ \sum_{j \in \mathcal{N}} \alpha_{j|i} (1 - T_{ij} + T_{ij} x_j) \right]^k
\end{aligned} \tag{S4}$$

by virtue of the multinomial theorem. The probability that a node of type $i$ has $\boldsymbol{k}$ nodes *directly* accessible (i.e., via an existing link) is the coefficient of $g_i(\boldsymbol{x})$ in front of $\prod_{j \in \mathcal{N}} x_j^{k_j}$; hence we say that $g_i(\boldsymbol{x})$ *generates* the joint degree distribution for nodes of type $i$. It is worth mentioning that disregarding the types of nodes [i.e., setting $x_j = x$ and $T_{ij} = T$ for all $i,j \in \mathcal{N}$] in Eq. (S4) yields

$$g(x) = \sum_k p_k [1 - T + Tx]^k \;, \tag{S5}$$

corresponding to the classical configuration model [2, 3]. In other words, even if distinguishing nodes into different types appears to force correlations in the way they are connected, our approach generates well-mixed networks that are structurally *indistinguishable* from networks generated with the configuration model and the same degree distribution.

To compute the size of the components requires to know the *excess* degree of nodes: the number of *new* nodes that can be reached from a node that has itself been reached via one of its links. The generating function associated with this distribution, $f_{li}(\boldsymbol{x})$, can readily be obtained from Eq. (S4) as

$$f_{li}(\boldsymbol{x}) := \left[ \frac{\partial g_i(\boldsymbol{x})}{\partial x_l} \right]_{\boldsymbol{x}=\boldsymbol{1}}^{-1} \left[ \frac{\partial g_i(\boldsymbol{x})}{\partial x_l} \right] = \sum_k \frac{k p_k}{\langle k \rangle} \left[ \sum_{j \in \mathcal{N}} \alpha_{j|i} (1 - T_{ij} + T_{ij} x_j) \right]^{k-1} . \tag{S6}$$

One striking result here is that $f_{li}(\boldsymbol{x})$ does not depend on the type of the node from which the node of type $i$ has been reached. This is a direct consequence of the multinomial form of Eq. (S2). Consequently, we define

$$f_i(\boldsymbol{x}) := f_{li}(\boldsymbol{x}) \tag{S7}$$

for the remaining of the document.

### A. Extensive (giant) component

With the generating function $g_i(\boldsymbol{x})$ and $f_i(\boldsymbol{x})$ in hand, we can already compute the size, composition and probability of the extensive, giant component in the limit of large size networks. Let us define $u_i$ as the probability

3that a neighbor of type $i$ does not lead to the giant component (i.e., regardless of the type of the node whose neighbor is of type $i$). Because links are created by randomly matching stubs, we expect the networks generated by our model to be locally tree-like in the limit of large networks for a great variety of degree distributions. Consequently, the probabilities $\{u_i\}_{i \in \mathcal{N}}$ can be obtained via a simple self-consistent argument: if the neighbor of type $i$ mentioned above does not lead to the extensive component, then neither should its other neighbors. Since the distribution of the number and of the type of these *other* neighbors is generated by $f_i(\boldsymbol{x})$, we directly have that the $\{u_i\}_{i \in \mathcal{N}}$ are the solutions of

$$u_i = f_i(\boldsymbol{u}) = \sum_k \frac{k p_k}{\langle k \rangle} \left[ \sum_{j \in \mathcal{N}} \alpha_{j|i} \left(1 - T_{ij} + T_{ij} u_j\right) \right]^{k-1} \tag{S8}$$

for every $i \in \mathcal{N}$. Following the same line of thoughts, the probability that a randomly chosen node of type $i$ leads to the giant component is simply the complement of the probability that neither of its neighbors leads to it, or in mathematical terms

$$P_i := 1 - g_i(\boldsymbol{u}) = 1 - \sum_k p_k \left[ \sum_{j \in \mathcal{N}} \alpha_{j|i} \left(1 - T_{ij} + T_{ij} u_j\right) \right]^{k} . \tag{S9}$$

The probability that any given node leads to the giant component is therefore

$$P := \sum_{i \in \mathcal{N}} w_i P_i = \sum_{i \in \mathcal{N}} w_i [1 - g_i(\boldsymbol{u})] = 1 - \sum_{i \in \mathcal{N}} w_i \sum_k p_k \left[ \sum_{j \in \mathcal{N}} \alpha_{j|i} \left(1 - T_{ij} + T_{ij} u_j\right) \right]^{k} . \tag{S10}$$

An important feature of asymmetric percolation on undirected networks – when links are undirected but the probability of existence of links is asymmetric (i.e., that $T_{ij} \neq T_{ji}$ from some $i$ and $j$) – is that it is equivalent to percolation on semi-directed networks. This implies that in general, the relative size of the giant component, $S$, does not equal its probability of existence, $P$ [1, 4]. In fact, while $P$ is computed by looking at the fraction of nodes that *lead* to the extensive component (i.e., extensive in-component), its relative size is obtained by looking at the fraction of nodes that can be *reached* from the formers (i.e., extensive out-component). Since links in our model are originally undirected, the relative size of the extensive component is obtained by simply switching the probability of existence of links; $T_{ij}$ becomes $T_{ji}$ for all $i, j \in \mathcal{N}$. Indeed, as explained in Sec. D, a neighbor of type $j$ can be reached from a node of type $i$ with probability $T_{ij} T_{ji} + T_{ij}(1 - T_{ji}) = T_{ij}$, while the same neighbor of type $j$ can reach the node of type $i$ with probability $T_{ij} T_{ji} + T_{ji}(1 - T_{ij}) = T_{ji}$. In other words, switching the probability of existence of links effectively corresponds to switching from the out-degree to the in-degree. Let us define $v_i$ as the probability that a node cannot be reached from the extensive component via a neighbor of type $i$, and using the same self-consistency argument as above, we find that it is the solution of

$$v_i = \sum_k \frac{k p_k}{\langle k \rangle} \left[ \sum_{j \in \mathcal{N}} \alpha_{j|i} \left(1 - T_{ji} + T_{ji} v_j\right) \right]^{k-1} \tag{S11}$$

for every $i \in \mathcal{N}$. Similarly, the probability that a node of type $i$ is not part of the extensive component, $S_i$, corresponds to the complement of the probability that none of its neighbors is part of it, or

$$S_i := 1 - \sum_k p_k \left[ \sum_{j \in \mathcal{N}} \alpha_{j|i} \left(1 - T_{ji} + T_{ji} v_j\right) \right]^{k} , \tag{S12}$$

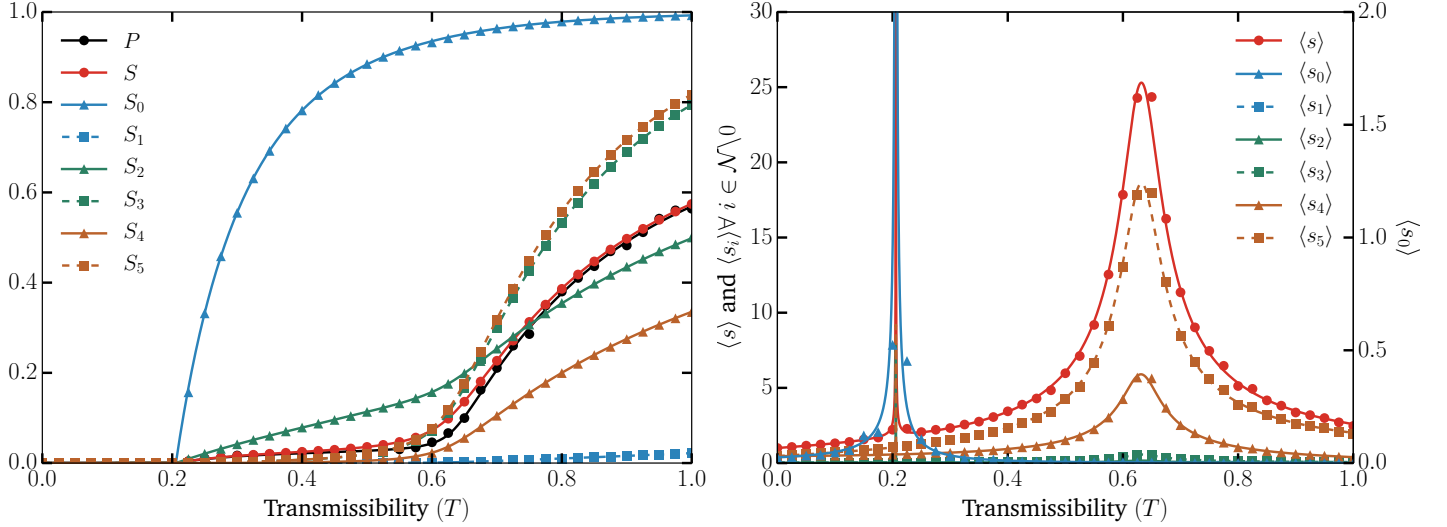

FIG. S1. **Validation of the analytical calculations.** The predictions (lines) of Eqs. (S8)–(S13) (left) and Eqs. (S21)–(S24) (right) are compared with the results obtained from numerical simulations (symbols). Note that $\langle s_0 \rangle$ is shown using a different axis to facilitate the comparison between the theoretical prediction and the results of the numerical simulations. The asymmetry parameter is $a = 10$; details on the other parameters and numerical simualtions are given in Secs. C and D.

and the relative size of the extensive component is simply the fraction of nodes that can be reached from other nodes in it, or

$$S := \sum_{i \in \mathcal{N}} w_i S_i = 1 - \sum_{i \in \mathcal{N}} w_i \sum_k p_k \left[ \sum_{j \in \mathcal{N}} \alpha_{j|i} \left(1 - T_{ji} + T_{ji} v_j\right) \right]^k . \tag{S13}$$

Since the right-hand side of Eqs. (S8) and (S11) are convex and normalized polynomials, the values of $\{u_i\}_{i \in \mathcal{N}}$ and $\{v_i\}_{i \in \mathcal{N}}$ can be obtained by iterating Eqs. (S8) and (S11) from an initial condition in $[0, 1)^{|\mathcal{N}|}$ until convergence. The predictions of Eqs. (S8)–(S13) are validated on Fig. S1.

### B. Non-extensive (small) components

The distribution of the composition of the small, non-extensive components can be computed in a similar fashion. Let us define the probability generating function (pgf) $H_i(\boldsymbol{x})$ whose coefficients correspond to the probability that a neighbor of type $i$ leads to a small component of a given composition (i.e., the number of nodes of type $j$ is given by the exponent of $x_j$). Invoking the same self-consistency argument as above, the pgfs are the solution of

$$H_i(\boldsymbol{x}) = x_i \sum_k \frac{k p_k}{\langle k \rangle} \left[ \sum_{j \in \mathcal{N}} \alpha_{j|i}[1 - T_{ij} + T_{ij} H_j(\boldsymbol{x})] \right]^{k-1} . \tag{S14}$$

where the extra $x_i$ has been added to account for the neighbor of type $i$ itself. Similarly, the small component that can be reached from a node of type $i$ is given by

$$K_i(\boldsymbol{x}) := x_i \sum_k p_k \left[ \sum_{j \in \mathcal{N}} \alpha_{j|i}[1 - T_{ij} + T_{ij} H_j(\boldsymbol{x})] \right]^k . \tag{S15}$$

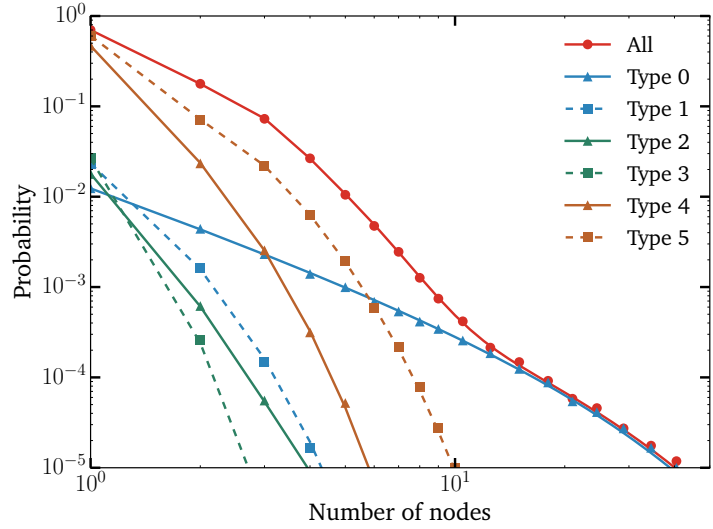

FIG. S2. **Validation of the analytical calculations.** The marginal distributions for the size of the small components predicted by Eqs. (S18)–(S20) (lines) are compared with the results obtained from numerical simulations (symbols). The asymmetry parameter is $a = 10$; details on the other parameters and numerical simualtions are given in Secs. C and D.

The distribution of the composition of the small components is

$$K(\boldsymbol{x}) := \sum_{i \in \mathcal{N}} w_i K_i(\boldsymbol{x}) = \sum_{i \in \mathcal{N}} w_i x_i \sum_k p_k \left[ \sum_{j \in \mathcal{N}} \alpha_{j|i}[1 - T_{ij} + T_{ij} H_j(\boldsymbol{x})] \right]^k . \tag{S16}$$

It is worth noting that whenever $S > 0$, the distribution generated by $K(\boldsymbol{x})$ is no longer normalized, $K(\mathbf{1}) < 1$, since there is a non-zero probability that a given node does not lead to a finite component. In fact, a comparison of Eqs. (S8)–(S10) with (S14)–(S16) yields the following relations

$$u_i = H_i(\mathbf{1}) \; ; \qquad P_i = 1 - K_i(\mathbf{1}) \; ; \qquad P = 1 - K(\mathbf{1}) \; . \tag{S17}$$

for every $i \in \mathcal{N}$.

Although the pgf $K(\boldsymbol{x})$ generates the joint distribution of the number of nodes of each type in small components, we are typically interested in its marginal distributions (i.e., the distribution of the number of nodes of a given type). To obtain the distribution of the number of nodes of a given type, say type $i$, we simply set $x_j = \delta_{ij} z$ for all $j \in \mathcal{N}$ in Eqs. (S14)–(S16). Similarly, if we are interested in the size of the components, that is the number of nodes regardless of their type, we simply set $x_j = z$ for all $j \in \mathcal{N}$. The marginal distribution for the number of nodes in a component, $P(s)$, is then

$$P(s) = \left[ \frac{1}{s!} \frac{\partial^s}{\partial z^s} \frac{K(z)}{K(1)} \right]_{z=0} . \tag{S18}$$

From Cauchy's integral formula, this last expression can be written in terms of a path integral in the complex plane

$$P(s) = \frac{1}{2i\pi K(1)} \oint_\Gamma \frac{K(z)}{z^{s+1}} dz \; , \tag{S19}$$

where $\Gamma$ is a closed circular path of radius $r$ centered at the origin and in which $K(z)$ is analytic. Substituting $z = re^{2i\pi\phi}$ and discretizing the integral in $s_{\max}$ equally spaced points, this last expression becomes

$$\begin{aligned} P(s) &= \frac{1}{2i\pi K(1)} \oint_\Gamma \frac{K(z)}{z^{s+1}} dz \\ &= \frac{1}{r^s K(1)} \int_0^1 K\big(re^{2i\pi\phi}\big) e^{-2i\pi s\phi} d\phi \\ &\simeq \frac{1}{s_{\max} r^s K(1)} \sum_{n=0}^{s_{\max}-1} K\big(re^{2i\pi n/s_{\max}}\big) e^{-2i\pi sn/s_{\max}} \; , \end{aligned} \qquad (S20)$$

where we identify the inverse discrete Fourier transform of the sequence $\{K(re^{2i\pi n/s_{\max}})\}$ with $n = 0, \ldots, s_{\max} - 1$. Although the exact form of $K(z)$ is unknown, it is nevertheless possible to evaluate the sequence $\{K(re^{2i\pi n/s_{\max}})\}$ by solving Eqs. (S14)–(S16) and then to quickly obtain $P(s)$ for $s < s_{\max}$ with a single pass of the FFT algorithm. This method is validated on Fig. S2.

The emergence of a critical cluster at the phase transition is associated with the divergence of the average size of the small components

$$\langle s \rangle = \sum_{i \in \mathcal{N}} \langle s_i \rangle \; , \qquad (S21)$$

which can be computed from the marginal distributions obtained with the method explained above. However, since $P(s) \sim s^{-3/2}$ at the phase transition [2], calculating $\langle s_i \rangle$ near the percolation threshold requires large values of $s_{\max}$ which make this method inefficient in practice. Luckily, there is another way to compute the average number of nodes in small components without having to compute the marginal distribution beforehand. Since the moments of a distribution can be obtained via a suitable differentiation of its associated pgf, the average number of nodes of type $i$ in small components can also be written as

$$\langle s_i \rangle = \frac{1}{K(\mathbf{1})} \left.\frac{dK(\boldsymbol{x})}{dx_i}\right|_{\boldsymbol{x}=\mathbf{1}} \; . \qquad (S22)$$

Substituting Eq. (S16) in this equation yields

$$\langle s_i \rangle = \frac{w_i g_i(\boldsymbol{u})}{K(1)} + \frac{\langle k \rangle}{K(1)} \sum_{j,l \in \mathcal{N}} w_j u_j \alpha_{l|j} T_{jl} \frac{\partial H_l(\mathbf{1})}{\partial x_i} \; , \qquad (S23)$$

where the equation for $\frac{\partial H_l(\mathbf{1})}{\partial x_i}$ is obtained via Eq. (S14)

$$\frac{\partial H_l(\mathbf{1})}{\partial x_i} = \delta_{il} f_l(\boldsymbol{u}) + \left[\sum_k \frac{k(k-1)p_k}{\langle k \rangle} \left(\sum_{j \in \mathcal{N}} \alpha_{j|l}[1 - T_{lj} + T_{lj}u_j]\right)^{k-2}\right] \times \left[\sum_{j \in \mathcal{N}} \alpha_{j|l} T_{lj} \frac{\partial H_j(\mathbf{1})}{\partial x_i}\right] \; , \qquad (S24)$$

for every $i, l \in \mathcal{N}$. Consequently, the average number of nodes of each type in small components can be obtained by solving Eq. (S24) for every combination of $i, l \in \mathcal{N}$ and then injecting the solutions in Eq. (S23). This approach is validated on Fig. S1.

### C. Choice of parameters

The proportion of each type of nodes, $\{w_i\}_{i \in \mathcal{N}}$ (see Table S1), used for the calculations are conservative yet realistic estimates extrapolated from the data provided in Refs. [5–7] with respect to the conclusions discussed in

| Description | Homo. males | Homo. females | Bi. males | Bi. females | Hetero. males | Hetero. females |
|---|---|---|---|---|---|---|
| $i$ | 0 | 1 | 2 | 3 | 4 | 5 |
| $w_i$ | 0.025 | 0.025 | 0.015 | 0.015 | 0.46 | 0.46 |

TABLE S1. Description of the node types and their relative proportion used in all calculations and numerical simulations.

the main text. The values chosen for the asymmetry parameter $a$ (i.e., 3 and 10), are also either quite conservative (3) or in line (10) with the naive estimates made from the infectious periods of males and females (see Introduction in the main text). Moreover, although there is evidence that the probability of HIV transmission differs by sexual act and insertive/receptor status (i.e., receptive anal sex could be as much as 20 times more likely to transmit HIV than insertive vaginal sex, see Ref. [8]), we are not aware of any strong evidence suggesting that the probability of transmission of ZIKV is different based on whether the *receptor* was male or female. We therefore opted for a conservative approach and chose to use the same probability of transmission regardless of the gender of the receiving partner (i.e., $T_{ij}$ does not depend on $j$), that is

$$T_{ij} = \begin{cases} T & \text{if } i \text{ is male} \\ T/a & \text{if } i \text{ is female} \end{cases} \tag{S25}$$

for every $j \in \mathcal{N}$. The use of fixed probabilities of transmission relies implicitly in the so-called i.i.d. assumption introduced in Ref. [3] which ignores, due to coarse-graining, some correlations and heterogeneity associated with the distribution of infectious periods. To understand the implications of the independent identically distributed (i.i.d.) assumption, let us consider an individual of type $i$ that is infectious for a period $\tau$ and that interacts with one of its neighbors of type $j$ at a rate $\beta$. The probability that this individual *never* infects its neighbor is

$$\lim_{\delta t \to 0} (1 - \beta \delta t)^{\tau/\delta t} = e^{-\beta \tau} . \tag{S26}$$

Assuming that $\tau$ and $\beta$ are i.i.d. random variables respectively drawn from the distributions $P_i(\tau)$ and $Q_{ij}(\beta)$, the probability that *any* individual of type $i$ *eventually* infects one of its neighbors of type $j$ is

$$T_{ij} = 1 - \iint e^{-\beta \tau} P_i(\tau) Q_{ij}(\beta) d\tau d\beta . \tag{S27}$$

The i.i.d. assumption therefore implies that each infection event is independent for the other infection events, even if these infection events all come from the same individual. However, although correlations in the infectious period at the individual level are not taken into account, they are taken into account at the gender level since the probability of transmission, $T_{ij}$, corresponds to the probability that a node of type $i$ will eventually infect its neighbor of type $j$ and, as such, is obtained by averaging over a distribution of infectious period, $P_i(\tau)$, that depends explicitly on $i$. Since the asymmetry in the probability of transmission between men and women is mainly due to men having longer infectious periods than women, the use of node types is expected to capture the relevant heterogeneity in the infectious period with regards to the double transition phenomenology. In other words, the i.i.d. assumption coupled with multiple types of nodes (i.e., the use of $T_{ij}$ instead of only one unique probability of transmission $T$) is sufficient to take into account that men are more likely to transmit ZIKV than women due to their longer infectious periods. Adapting the frameworks developed in Refs. [9, 10] to multiple types of nodes would get rid of the i.i.d. asumption but is not expected to alter the qualitative aspect of the double transition phenomenology.

The values of $\{\alpha_{j|i}\}_{i,j \in \mathcal{N}}$ are chosen to respect as much as possible the proportion of each type of nodes, $\{w_i\}_{i \in \mathcal{N}}$ (see Table S1), given the constraints of Eqs. (S1). Taking the nodes of type 2 as a starting point, we set

$$\alpha_{0|2} = \frac{w_0}{w_0 + w_2 + w_3 + w_5} \; ; \qquad \alpha_{2|2} = \frac{w_2}{w_0 + w_2 + w_3 + w_5}$$
$$\alpha_{3|2} = \frac{w_3}{w_0 + w_2 + w_3 + w_5} \; ; \qquad \alpha_{5|2} = \frac{w_5}{w_0 + w_2 + w_3 + w_5} \; ,$$

| i \ j | 0 | 1 | 2 | 3 | 4 | 5 |
|---|---|---|---|---|---|---|
| 0 | 0.971 | 0.000 | 0.029 | 0.000 | 0.000 | 0.000 |
| 1 | 0.000 | 0.971 | 0.000 | 0.029 | 0.000 | 0.000 |
| 2 | 0.049 | 0.000 | 0.029 | 0.029 | 0.000 | 0.893 |
| 3 | 0.000 | 0.049 | 0.029 | 0.029 | 0.893 | 0.000 |
| 4 | 0.000 | 0.000 | 0.000 | 0.029 | 0.000 | 0.971 |
| 5 | 0.000 | 0.000 | 0.029 | 0.000 | 0.971 | 0.000 |

TABLE S2. Values of $\alpha_{j|i}$ used in all calculations and numerical simulations.

from which we can readily obtain the following from Eqs. (S1)

$$\alpha_{2|0} = \frac{w_2 \alpha_{0|2}}{w_0} \ ; \quad \alpha_{2|3} = \frac{w_2 \alpha_{3|2}}{w_3} \ ; \quad \alpha_{2|5} = \frac{w_2 \alpha_{5|2}}{w_5} \ : \quad \alpha_{0|0} = 1 - \alpha_{2|0}$$

$$\alpha_{4|5} = 1 - \alpha_{2|5} \ ; \quad \alpha_{5|4} = \frac{w_5 \alpha_{4|5}}{w_4} \ ; \quad \alpha_{3|4} = 1 - \alpha_{5|4} \ ; \quad \alpha_{4|3} = \frac{w_4 \alpha_{3|4}}{w_3} \ .$$

Taking a similar point of view for nodes of type 3, but taking into account that $\alpha_{2|3}$ and $\alpha_{4|3}$ are already known, we set

$$\alpha_{1|3} = \frac{[1 - (\alpha_{2|3} + \alpha_{4|3})]w_1}{w_1 + w_3} \ ; \quad \alpha_{3|3} = \frac{[1 - (\alpha_{2|3} + \alpha_{4|3})]w_3}{w_1 + w_3} \ ,$$

from which we finally obtain

$$\alpha_{3|1} = \frac{w_3 \alpha_{1|3}}{w_1} \ ; \quad \alpha_{1|1} = 1 - \alpha_{3|1} \ .$$

All other probabilities are set to zero. The values obtained using the parameters given in Table S1 are given in Table S2.

### D. Details on the numerical simulations

The analytical predictions of the model are validated by comparing them with results obtained from numerical simulations in which networks are generated according to the stub matching scheme of Ref. [1]. More precisely, $10^6$ nodes are first assigned a number of stubs, corresponding to their degree, drawn from a Poisson distribution

$$p_k = \frac{e^{-\langle k \rangle} \langle k \rangle^k}{k!} \tag{S28}$$

where $k \geq 0$ and with an average degree $\langle k \rangle = 5$. Stubs are then paired up to form links as follows.

1. The stubs assigned to each node are gathered in six lists of unmatched stubs (one list for each node type).
2. With probability

$$R_{ij} = \frac{w_i \alpha_{j|i}}{1 + \delta_{ij}} \left[ \sum_{m \geq n \in \mathcal{N}} \frac{w_m \alpha_{n|m}}{1 + \delta_{mn}} \right]^{-1}, \tag{S29}$$

one stub attached to a node of type $i$ and one stub attached to a node of type $j$ are randomly chosen and are matched to form a link. These stubs are then removed from their respective list.

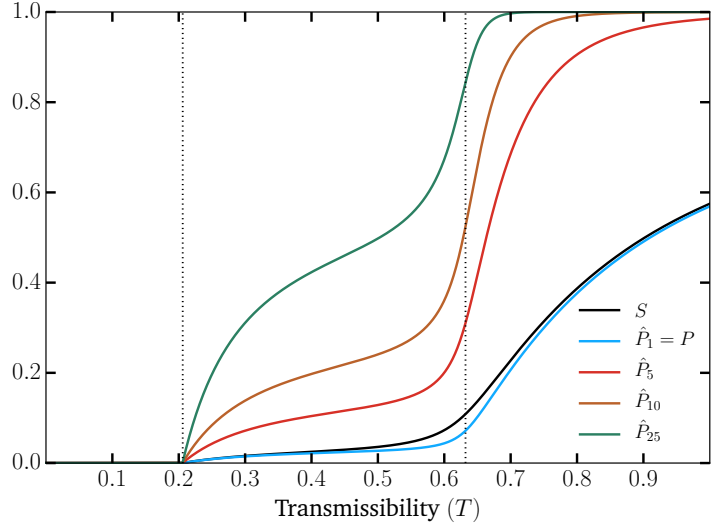

FIG. S3. **Effect of multiple seeds.** Behavior of Eq. (S30) for various values of $N$ using the same parameters as in Fig. S1. The two vertical dotted lines show the thresholds for the the two transitions as introduced in the main text.

3. Step 2 is repeated until every list is *effectively* empty (i.e., some stubs may remained unmatched due to statistical fluctuations but their number is typically small enough such that the observables measured on the generated networks are not affected).

The generated *undirected* network is then transformed into a *semi-directed* network as follows. A link between a node of type $i$ and a node of type $j$

- remains undirected with probability $T_{ij}T_{ji}$;
- becomes directed from the node of type $i$ to the node of type $j$ ($i \to j$) with probability $T_{ij}(1 - T_{ji})$;
- becomes directed from the node of type $j$ to the node of type $i$ ($j \to i$) with probability $T_{ji}(1 - T_{ij})$;
- is removed with probability $(1 - T_{ij})(1 - T_{ji})$.

This procedure is repeated for every link. The probability that a node of type $i$ starts an epidemic, $P_i$, then corresponds to the fraction of nodes of type $i$ that are in the extensive in-component, and the probability that a node of type $i$ is infected during an epidemic, $S_i$, corresponds to the fraction of nodes of type $i$ that are in the extensive out-component. Similarly, the average composition of outbreaks, $\langle s_i \rangle \ \forall i \in \mathcal{N}$, are obtained by looking at the composition of non-extensive out-components.

Since the observables measured on Figs. S1 and S2 are self-averaging properties, the number of semi-directed networks over which they were averaged was chosen to be large enough so that the stochastic fluctuations were reduced to a satisfactory level (about 5000 for each value of $T$).

### E. Effect of multiple seeds

The effect of having multiple seeds due to different individuals being infected by mosquitoes can be investigated using our formalism. In Sec. A, we define $P$ as the probability that a random individual sparks an epidemic of size $S$. If we assume that $N$ such individuals have been independently infected by mosquitoes (i.e., they are sufficiently far apart in the contact network), the probability that at least one of them sparks an epidemic is

$$\hat{P}_N = 1 - (1 - P)^N \ . \tag{S30}$$

9

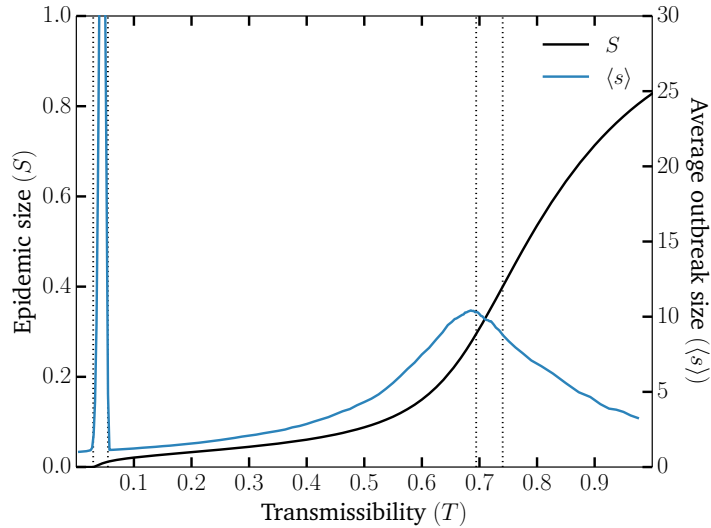

FIG. S4. **Simulations using realistic data.** The size of the extensive component, $S$, and the average size of microscopic outbreaks, $\langle s \rangle$ are shown for the random network ensemble defined in Sec. F. These results confirm the general phenomenology presented in the main text as well as the qualitative accuracy of the naive estimates of the upper and lower bounds of the thresholds (black dotted vertical lines). Note that the distribution of the size of the small components follows a power law on many orders of magnitude at the second transition. This obfuscates the position of the peak in practice for finite-size networks. Based on Fig. 2 in the main text, we expect the peak to be slightly at the right of the line at $T \simeq 0.69$.

Moreover, since a fraction $1-P$ of all seeds are expected to *not* spark a macroscopic epidemic, $N$ seeds are expected to infect $(1-P)N\langle s \rangle$ individuals in microscopic outbreaks. As $N$ increases, we therefore expect a higher probability of observing a macroscopic epidemic as well as more individuals infected in microscopic outbreaks. Figure S3 shows the behavior of the quantity $\hat{P}_N$ for the parameters used in Fig. S1. We see that although multiple seeds tend to smooth the transition between the two thresholds with respect to the probability of sparking an epidemic, a similar phenomenology is observed. More importantly, as shown in Ref. [11], the size of the epidemic is unaffected by a finite number of seeds in the limit of large network size. In other words, having multiple seeds increases the likelihood of reaching the extensive out-component, but it does not change its size. While this only maps to vector transmission once a steady-state is achieved, it does suggest that the general phenomenology presented in the main text is valid with or without vector transmission.

### F. Simulations using realistic data

To further support our results, we simulated ZIKV transmission dynamics on a random network of sexual contacts where the distribution of contacts are drawn from Ref. [5] (for homosexual contacts) and Ref. [7] (for heterosexual contacts) but only using individuals with more than one sexual partner (as a rough approximation of the sexually active population). A population of $10^6$ individuals is assumed to be equally parted between men and women, of which 5% are homosexuals, 3% are bisexuals and 92% are heterosexuals. We assume very few individuals are infected by the mosquito vector, but they can cause outbreaks through sexual transmission where men transmit with probability $T$ and women with probability $T/a$. Using $a = 2$, Fig. S4 shows that increasing heterogeneity in the distribution of the degrees does not alter the phenomenology presented in the main text. Moreover, we see that the "naive" estimates described in the caption of Fig. 2 in the main text remain good

indicators for the location of the two epidemic thresholds.


\end{document}